\documentclass[12pt]{article}
\usepackage{amsmath}
\usepackage{amsfonts}
\usepackage{amssymb}
\usepackage[usenames]{color}
\usepackage{amsthm}

\def\p{\partial}
\def\e{\mathrm{e}}

\theoremstyle{remark}

\newcommand{\dbar}{\bar{\partial}}

\newcommand{\be}{\begin{equation}}
\newcommand{\ee}{\end{equation}}
\newcommand{\bea}{\begin{eqnarray}}
\newcommand{\eea}{\end{eqnarray}}
\newcommand{\beaa}{\begin{eqnarray*}}
\newcommand{\eeaa}{\end{eqnarray*}}

\newcommand{\nn}{\nonumber}
\renewcommand{\d}{\mathrm{d}}

\usepackage{authblk}
\begin{document}
\title
{The Orlov-Schulman symmetries of the Manakov-Santini hierarchy
}
\author{L.V. Bogdanov\thanks{leonid@itp.ac.ru}}
\affil{Landau Institute for Theoretical Physics RAS}
\date{}
\maketitle
\begin{abstract}
We construct the Orlov-Schulman symmetries for the Manakov-Santini (MS) hierarchy.
We give  an explicit proof of compatibility of additional symmetries with the basic flows of the MS hierarchy, and  consider several simple examples, including the Galilean transformation and 
scalings. We also present a picture of the Orlov-Schulman symmetries in terms of the dressing
scheme based on the Riemann-Hilbert problem. 
\end{abstract}
\section{Introduction}
In this work we construct the Orlov-Schulman symmetries for the Manakov-Santini (MS) hierarchy.
The Orlov-Schulman symmetries were introduced 
in the context of the KP (Kadomtsev-Petviashvili)  hierarchy 
\cite{OS,Orlov88,GrinOrlov89}, they
form a Lie algebra of symmetries of the KP hierarchy, commuting with all the flows of the
hierarchy but not between themselves. The Orlov-Schulman symmetries are now widely used
in the applications of the KP hierarchy to the field theory, matrix integrals, string equations etc.
For the dispersionless limit of KP hierarchy (the dKP hierarchy) these symmetries were constructed
by Takasaki and Takebe \cite{Takasaki92}.

The MS system was introduced rather recently as a generalisation 
of the dKP equation \cite{MS06,MS07,MS08}.
It was discovered \cite{DFK15} that this system possesses a deep geometric meaning,
describing a general local form of Einstein-Weyl spaces in (2+1) dimensions.
The dKP hierarchy corresponds to the group of volume-preserving diffeomorphisms Sdiff(2)
with the symmetries represented by Hamiltonian vector fields,
while the MS hierarchy \cite{BDM07,LVB09} is defined by the group of diffeomorphisms Diff(2)
with the symmetries represented by arbitrary vector fields. The MS system belongs to the class
of multidimensional dispersionless integrable systems having in general no direct analogue in the
class of class of integrable systems with dispersion \cite{BDM07,LVB09}. Thus the Orlov-Schulman
symmetries for this class cannot be obtained using the dispersionless limit.
However, the construction developed by Takasaki for the dKP hierarchy and Hamiltonian
vector fields (Poisson algebra) can be generalised to the case of general vector fields
and associated  multidimensional dispersionless integrable hierarchies.
Here we construct the Orlov-Schulman symmetries for the Manakov-Santini hierarchy case.
Having in mind the existing gap in the literature concerning this topic, we present it in some detail,
giving an explicit proof of compatibility of additional symmetries with the basic flows of the MS hierarchy, and  consider several simple examples.
We also present a picture of the Orlov-Schulman symmetries in terms of the dressing
scheme based on the Riemann-Hilbert problem \cite{BDM07,LVB09}.
\section{The MS hierarchy}
The Manakov-Santini hierarchy is defined by the Lax-Sato equations 
\cite{BDM07,LVB09}
\bea
&&
\frac{\partial}{\partial t_n}\begin{pmatrix}
L\\
M
\end{pmatrix}=
\left(\left(
\frac{ L^n L_p}{\{L,M\}}\right)_+
{\partial_x}
-\left(\frac{ L^n L_x}{\{L,M\}}\right)_+
{\partial_p}\right)\begin{pmatrix}
L\\
M
\end{pmatrix},
\label{genSato1}
\eea 
where $L$, $M$, corresponding to the Lax and Orlov functions of 
the dispersionless KP hierarchy,
are formal series
\bea
&&
L=p+\sum_{n=1}^\infty u_n(\mathbf{t})p^{-n},
\label{form01}
\\&&
M=\sum_{n=0}^\infty t_n L^{n} + \sum_{n=1}^\infty v_n(\mathbf{t})p^{-n}
\label{form1}
,
\eea
and $x=t_0$, $(\sum_{-\infty}^{\infty}u_n p^n)_+
=\sum_{n=0}^{\infty}u_n p^n$, $\{L,M\}=L_pM_x-L_xM_p$.
By construction, due to the structure of the series
(\ref{form01}), (\ref{form1}), the coefficients of vector fields 
in Lax-Sato equations (\ref{genSato1}) are polynomials in $p$
($\{L;M\}$ is a formal series in $p$ with the
first terms $1+v_x p^{-1} +\dots$).

A more standard choice of times for the dKP hierarchy corresponds to 
$M$=${\sum_{n=0}^\infty (n+1)t_n L^{n}}+\dots$, it differs 
by the scaling of times.

Lax-Sato equations for the first two flows of the hierarchy (\ref{genSato1})
\bea
&&
\partial_y
\begin{pmatrix}
L\\
M
\end{pmatrix}
=((p-v_{x})\partial_x - u_{x}\partial_p)
\begin{pmatrix}
L\\
M
\end{pmatrix}
\label{MSLax1},
\\
&&
\partial_t
\begin{pmatrix}
L\\
M
\end{pmatrix}
=((p^2-v_{x}p+u -v_{y})\partial_x
-(u_{x}p+u_{y})\partial_p)
\begin{pmatrix}
L\\
M
\end{pmatrix}
\label{MSLax2},
\eea
where $u=u_1$, $v=v_1$, $x=t_0$, $y=t_1$, $t=t_2$,
correspond to the Lax pair  of the Manakov-Santini system 
\bea
u_{xt} &=& u_{yy}+(uu_x)_x+v_xu_{xy}-u_{xx}v_y,
\nn\\
v_{xt} &=& v_{yy}+uv_{xx}+v_xv_{xy}-v_{xx}v_y,
\label{MSeq}
\eea
The MS hierarchy is deeply connected with a special kind of Riemann-Hilbert problem, 
which clarifies the structure of the hierarchy and
leads to a generating relation representing the hierarchy in a compact form. However, we prefer
to first consider the hierarchy in the form (\ref{genSato1}) per se, as equations defining the dynamics
for formal series
$L$, $M$ with the coefficients depending on $x$. The hierarchy (\ref{genSato1}) is relatively new,
thus we hope some technical details will be helpful
and fill the gap in the existing literature. The dressing scheme based on the Riemann-Hilbert problem
is presented below.
\subsection*{Compatibility of the flows of the hierarchy}
Before introducing the Orlov-Schulman symmetries for the Manakov-Santini hierarchy,
we will demonstrate the compatibility of the hierarchy flows. Technically the proof is similar 
to the standard dispersionless KP (dKP) hierarchy case, but there are some peculiarities connected
with the fact that vector fields are not Hamiltonian as in the dKP hierarchy case. For simplicity
we will usually suggest that only finite number of times of the hierarchy is not equal to zero.
In the general multidimensional case the proof was sketched in the aricle \cite{BDM07}. We believe it will be useful for the reader to have a detailed calculation for the MS hierarchy, since it will help to understand the compatibility of  the Orlov-Schulman symmetries with the basic hierarchy.
To make the calculations more transparent, we write down Lax-Sato equations (\ref{genSato1})
in a concise form
\bea
&&
\frac{\partial}{\partial t_n}\begin{pmatrix}
L\\
M
\end{pmatrix}=
\hat V^n_+
\begin{pmatrix}
L\\
M
\end{pmatrix},
\label{genSato1v}
\eea
where $\hat V^n_+$ are vector fields defined by (\ref{genSato1}). We will also use `minus'  vector field
 $\hat V^n_-$ and vector field  $\hat V^n$, for which the coefficients are taken respectively with `minus' projection (projection to negative powers of $p$) and without projection.
Compatibility of two flows corresponding to some $t_n$, $t_m$ is expressed by the condition
\bea
\frac{\partial}{\partial t_n}
\left(
\hat V^m_+
\begin{pmatrix}
L\\
M
\end{pmatrix}
\right)
=
\frac{\partial}{\partial t_m}
\left(
\hat V^n_+
\begin{pmatrix}
L\\
M
\end{pmatrix}
\right)
\label{comp0}
\eea
In other words, the differential form
\beaa
\omega=\sum_{n=1}^\infty
\left(
\hat V^n_+
\begin{pmatrix}
L\\
M
\end{pmatrix}
\right)d t_n
\eeaa
should be closed.  Applying equations of the hierarchy (\ref{genSato1v}), we obtain
\bea
&&
\frac{\partial}{\partial t_n}
\left(
\hat V^m_+
\begin{pmatrix}
L\\
M
\end{pmatrix}
\right)
-
\frac{\partial}{\partial t_m}
\left(
\hat V^n_+
\begin{pmatrix}
L\\
M
\end{pmatrix}
\right)
\nn\\&&\qquad
=
\left(
((\p_n\hat V^m_+)- (\p_m\hat V^n_+)
+[\hat V^m_+;\hat V^n_+])
\begin{pmatrix}
L\\
M
\end{pmatrix}
\right)
\label{comp1}
\eea
Using an identity
\beaa
\hat V^n
\begin{pmatrix}
L\\
M
\end{pmatrix}=
\begin{pmatrix}
0\\
L^n
\end{pmatrix},
\eeaa
we can rewrite Lax-Sato equations (\ref{genSato1v}) in terms of `minus' vector fields,
\bea
&&
\frac{\partial}{\partial t_n}\begin{pmatrix}
L\\
M
\end{pmatrix}=
-\hat V^n_-
\begin{pmatrix}
L\\
M
\end{pmatrix}
+
\begin{pmatrix}
0\\
L^n
\end{pmatrix}.
\label{genSato1v-}
\eea
Performing the calculations similar to the `plus' case, we obtain
\beaa
&&
\frac{\partial}{\partial t_n}
\left(
-\hat V^m_-
\begin{pmatrix}
L\\
M
\end{pmatrix}
+
\begin{pmatrix}
0\\
L^m
\end{pmatrix}
\right)
-
\frac{\partial}{\partial t_m}
\left(-
\hat V^n_-
\begin{pmatrix}
L\\
M
\end{pmatrix}
+
\begin{pmatrix}
0\\
L^n
\end{pmatrix}
\right)
\\ &&\qquad\qquad
=
\left(
((\p_m\hat V^n_-)- (\p_n\hat V^m_-)
+[\hat V^m_-;\hat V^n_-])
\begin{pmatrix}
L\\
M
\end{pmatrix}
\right).
\eeaa
Comparing `plus' and `minus'  expressions for the same object, we get the equality
\beaa
&&
\left(
((\p_n\hat V^m_+)- (\p_m\hat V^n_+)
+[\hat V^m_+;\hat V^n_+])
\begin{pmatrix}
L\\
M
\end{pmatrix}
\right)
\\ 
&&
\qquad
=
\left(
((\p_m\hat V^n_-)- (\p_n\hat V^m_-)
+[\hat V^m_-;\hat V^n_-])
\begin{pmatrix}
L\\
M
\end{pmatrix}
\right).
\eeaa
Taking into account that $\{L;M\}\neq 0$ ($\{L;M\}$ is a formal series in $p$ with the
first terms $1+v_x p^{-1} +\dots$), and vector fields are two-dimensional, 
we come to the conclusion that
\beaa
&&
((\p_n\hat V^m_+)- (\p_m\hat V^n_+)
+[\hat V^m_+;\hat V^n_+])
-
((\p_n\hat V^m_-)- (\p_m\hat V^n_-)
+[\hat V^m_-;\hat V^n_-])
=0.
\eeaa
For nonnegative powers of $p$ (`plus' projection), we obtain
\beaa
&&
((\p_n\hat V^m_+)- (\p_m\hat V^n_+)
+[\hat V^m_+;\hat V^n_+])=0,
\eeaa
and formula (\ref{comp1}) implies compatibility condition (\ref{comp0}), Q.E.D.
\section{The Orlov-Schulman symmetries}
We introduce the Orlov-Schulman symmetries through some special `minus' vector fields,
similar to Takasaki and Takebe construction for the dKP hierarchy case \cite{Takasaki92},
\bea
&&
\frac{\partial}{\partial \tau}\begin{pmatrix}
L\\
M
\end{pmatrix}=
\left(-\left(
\frac{F^1 F^2_p}{\{L,M\}}\right)_-
{\partial_x}
+\left(\frac{F^1 F^2_x}{\{L,M\}}\right)_-
{\partial_p}\right)\begin{pmatrix}
L\\
M
\end{pmatrix},
\label{OS}
\eea 
where $F^1=F^1(L,M)$, $F^2=F^2(L,M)$ are arbitrary wave functions of the linear equations of the hierarchy,
\bea
&&
\frac{\partial}{\partial t_n}
F^i=
\left(\left(
\frac{ L^n L_p}{\{L,M\}}\right)_+
{\partial_x}
-\left(\frac{ L^n L_x}{\{L,M\}}\right)_+
{\partial_p}\right)
F^i.
\label{Linear}
\eea 
The origin of these flows will be 
more clear in the context of the Riemann-Hilbert problem based dressing scheme, 
which we consider below.
As we will demonstrate now, each of the flows (\ref{OS}) commutes with all the basic flows of the hierarchy
(\ref{genSato1}). However, in general the Orlov-Schulman flows 
do not commute between themselves,
and it is convenient to consider them in the context of infinitesimal symmetries of the hierarchy,
then formulae (\ref{OS}) give generators of infinitesimal symmetries.

Using an identity
\bea
\left(\left(
\frac{F^1F^2_p}{\{L,M\}}\right)
{\partial_x}
-\left(\frac{F^1F^2_x}{\{L,M\}}\right)
{\partial_p}\right)
\begin{pmatrix}
L\\
M
\end{pmatrix}
=
\begin{pmatrix}
-F^1 F^2_M\\
F^1 F^2_L
\end{pmatrix},
\label{idOS}
\eea
we can rewrite (\ref{OS}) in terms of `plus' vector fields,
\bea
&&
\frac{\partial}{\partial \tau}
\begin{pmatrix}
L\\
M
\end{pmatrix}=
\left(\left(
\frac{F^1F^2_p}{\{L,M\}}\right)_+
{\partial_x}
-\left(\frac{F^1F^2_x}{\{L,M\}}\right)_+
{\partial_p}\right)
\begin{pmatrix}
L\\
M
\end{pmatrix}
+
\begin{pmatrix}
F^1F^2_M\\
-F^1 F^2_L
\end{pmatrix}\qquad
\label{OS+}
\eea 
Below we will use a concise notation for the flows (\ref{OS}),
\beaa
\frac{\partial}{\partial \tau}\begin{pmatrix}
L\\
M
\end{pmatrix}=
-\hat U_-
\begin{pmatrix}
L\\
M
\end{pmatrix},
\eeaa
and identity (\ref{idOS}) reads
\beaa
\hat U
\begin{pmatrix}
L\\
M
\end{pmatrix}
=
\begin{pmatrix}
-F^1 F^2_M\\
F^1 F^2_L
\end{pmatrix}.
\eeaa
\subsection*{Compatibility with the hierarchy}
Writing down the flows of the hierarchy and Orlov-Schulman symmetry (\ref{OS}) in terms of 
`plus' vector fields,
\beaa
\frac{\partial}{\partial t_n}\begin{pmatrix}
L\\
M
\end{pmatrix}=
\hat V^n_+
\begin{pmatrix}
L\\
M
\end{pmatrix},
\quad
\frac{\partial}{\partial \tau}\begin{pmatrix}
L\\
M
\end{pmatrix}=
\hat U_+
\begin{pmatrix}
L\\
M
\end{pmatrix}
+
\begin{pmatrix}
F^1 F^2_M\\
-F^1 F^2_L
\end{pmatrix},
\eeaa
we get a compatibility condition in the form
\beaa
\frac{\partial}{\partial t_n}
\left(
\hat U_+
\begin{pmatrix}
L\\
M
\end{pmatrix} + 
\begin{pmatrix}
F^1F^2_M\\
-F^1F^2_L
\end{pmatrix}
\right)
=
\frac{\partial}{\partial \tau}
\left(
\hat V^n_+
\begin{pmatrix}
L\\
M
\end{pmatrix}
\right).
\eeaa
After simple transformations, 
\bea
&&
\frac{\partial}{\partial t_n}
\left(
\hat U_+
\begin{pmatrix}
L\\
M
\end{pmatrix} + 
\begin{pmatrix}
F^1 F^2_M\\
-F^1 F^2_L
\end{pmatrix}
\right)
-
\frac{\partial}{\partial \tau}
\left(
\hat V^n_+
\begin{pmatrix}
L\\
M
\end{pmatrix}
\right)
\nn\\ 
&&\qquad\qquad
=((\p_n\hat U_+)- (\p_\tau\hat V^n_+)
+[\hat U_+;\hat V^n_+])
\begin{pmatrix}
L\\
M
\end{pmatrix}.
\label{tr1}
\eea
On the other hand,
in terms of `minus' vector fields
\beaa
\frac{\partial}{\partial t_n}\begin{pmatrix}
L\\
M
\end{pmatrix}=
-\hat V^n_-
\begin{pmatrix}
L\\
M
\end{pmatrix}
+
\begin{pmatrix}
0\\
L^n
\end{pmatrix}
,
\quad
\frac{\partial}{\partial \tau}\begin{pmatrix}
L\\
M
\end{pmatrix}
=
-\hat U_-
\begin{pmatrix}
L\\
M
\end{pmatrix},
\eeaa
and the same compatibility condition is written in the form
\beaa
\frac{\partial}{\partial t_n}
\left(-
\hat U_-
\begin{pmatrix}
L\\
M
\end{pmatrix} 
\right)
=
\frac{\partial}{\partial \tau}
\left(-
\hat V^n_-
\begin{pmatrix}
L\\
M
\end{pmatrix}
+
\begin{pmatrix}
0\\
L^n
\end{pmatrix}
\right).
\eeaa
Using the equality
\beaa
&&
\frac{\partial}{\partial t_n}
\left(-
\hat U_-
\begin{pmatrix}
L\\
M
\end{pmatrix}  
\right)
-
\frac{\partial}{\partial \tau}
\left(-
\hat V^n_-
\begin{pmatrix}
L\\
M
\end{pmatrix}
+
\begin{pmatrix}
0\\
L^n
\end{pmatrix}
\right)
\\ &&\qquad\qquad
=
((\p_\tau\hat V^n_-)-(\p_n\hat U_-)
+[\hat U_-;\hat V^n_-])
\begin{pmatrix}
L\\
M
\end{pmatrix}
\eeaa
and comparing with another expression for the same object (\ref{tr1}),
we come to the conclusion that
\beaa
&&
\bigl(((\p_n\hat U_+)- (\p_\tau\hat V^n_+)
+[\hat U_+;\hat V^n_+])
\\ &&\qquad
-
((\p_\tau\hat V^n_-)-(\p_n\hat U_-)
+[\hat U_-;\hat V^n_-])\bigr)
\begin{pmatrix}
L\\
M
\end{pmatrix}
=0,
\eeaa
which implies, in complete analogy with the previous subsection,
that  the  vector field  in the l.h.s. and both its `plus' and `minus' projections are
equal to zero, thus proving compatibility 
of an Orlov-Schulman flow (\ref{OS}) with the MS hierarchy
(\ref{genSato1}), Q.E.D.
\section{Examples of the Orlov-Schulman symmetries}
We will consider some simplest examples of Orlov-Schulman symmetries (\ref{OS}),
corresponding to a special choice  of functions $F^1$, $F^2$.
\subsection{The Galilean transformation} 
Let us take $F^1=1$, $F^2=M$. The  Orlov-Schulman flow written in terms 
of `plus' vector fields (\ref{OS+}) reads
\bea
&&
\frac{\partial}{\partial \tau}
\begin{pmatrix}
L\\
M
\end{pmatrix}=
\left
(\left(
\frac{M_p}{\{L,M\}}\right)_+
{\partial_x}
-\left(\frac{M_x}{\{L,M\}}\right)_+
{\partial_p}\right)
\begin{pmatrix}
L\\
M
\end{pmatrix}
+
\begin{pmatrix}
1\\
0
\end{pmatrix}.
\label{OS+1}
\eea 
Using the identities
\beaa
\left(
\frac{M_p}{\{L,M\}}\right)_+
=
\left(
\frac{\p_p\sum t_n L^n}{\{L,M\}}\right)_+,
\quad
\left(
\frac{M_x}{\{L,M\}}\right)_+
=
\left(
\frac{\p_x\sum t_n L^n}{\{L,M\}}\right)_+,
\eeaa
we can rewrite the flow (\ref{OS+1}) as
\beaa
&&
\frac{\partial}{\partial \tau}L=1 +\left(-\p_p + \sum_{n} (n+1) t_{n+1} \p_{n}\right)L  ,
\\
&&
\frac{\partial}{\partial \tau}M=\left(-\p_p + \sum_n (n+1) t_{n+1} \p_{n}\right)M.
\eeaa
Taking the times $t_n$ with $n>N$ equal to zero, we get
\beaa
\frac{\partial}{\partial \tau}\sum_{n=0}^N t_n L^n
=\left(-\p_p + \sum^{N-1}_{k=0} (k+1) t_{k+1} \p_{k}\right)\sum_{n=0}^N t_n L^n
\eeaa
Let us restrict ourselves to the first three variables $x=t_0$, $y=t_1$, $t=t_2$, corresponding
to MS system (\ref{MSeq}). In this case
\beaa
&&
\frac{\partial}{\partial \tau}L=1 +\left(-\p_p + y\p_x+2t\p_y \right)L  ,
\\
&&
\frac{\partial}{\partial \tau}M=\left(-\p_p +  y\p_x+2t\p_y \right)M
\eeaa
Taking $p$ expansion (see (\ref{form01}),(\ref{form1})), for 
the functions $u$ and $v$ we get
\beaa
&&
\p_\tau u=\left(y\p_x+2t\p_y \right)u ,
\\
&&
\p_\tau v=\left(y\p_x+2t\p_y \right)v,
\eeaa
thus
\bea
u(\tau)=u(x+\tau y + \tau^2 t, y + 2\tau t, t),
\nn\\
v(\tau)=u(x+\tau y + \tau^2 t, y + 2\tau t, t).
\label{Galilean}
\eea
This symmetry represents a well-known Galilean transformation, 
it is compatible with the reductions
$v=0$ and $u=0$, corresponding respectively to the dKP equation and the Pavlov-Mikhalev equation.
It is easy to check that transformation (\ref{Galilean}) is a symmetry of MS system (\ref{MSeq}) directly.
\subsection{Dispersionless scaling} 
Let us take $F^1=M$, $F^2=L$. The  Orlov-Schulman flow written in terms 
of `plus' vector fields (\ref{OS+}) reads
\bea
&&
\frac{\partial}{\partial \tau}
\begin{pmatrix}
L\\
M
\end{pmatrix}=
\left
(\left(
\frac{ML_p}{\{L,M\}}\right)_+
{\partial_x}
-\left(\frac{ML_x}{\{L,M\}}\right)_+
{\partial_p}\right)
\begin{pmatrix}
L\\
M
\end{pmatrix}
+
\begin{pmatrix}
0\\
-M
\end{pmatrix}
\quad
\label{OS+2}
\eea 
Using the identities
\beaa
\left(
\frac{ML_p}{\{L,M\}}\right)_+=\left(
\frac{L_p\sum t_n L^n}{\{L,M\}}\right)_+,
\quad
\left(
\frac{ML_x}{\{L,M\}}\right)_+=\left(
\frac{L_x\sum t_n L^n}{\{L,M\}}\right)_+,
\eeaa
we rewrite flow (\ref{OS+2}) as
\beaa
\frac{\partial}{\partial \tau}L=\sum_{n} t_{n} \p_{n}L  ,
\quad
\frac{\partial}{\partial \tau}M=-M +\sum_n t_{n} \p_{n}M,
\eeaa
and for the functions $u$, $v$ we get
\beaa
\p_\tau u=\sum_{n} t_{n} \p_{n}u ,
\quad
\p_\tau v=-v + \sum_{n} t_{n} \p_{n}v.
\eeaa
Thus, in explicit form
\bea
&&
u(\tau)=u(\e^\tau x,\e^\tau y , \e^\tau t,\dots, \e^\tau t_n,\dots),
\nn\\
&&
v(\tau)=\e^{-\tau}v(\e^\tau x,\e^\tau y , \e^\tau t,\dots, \e^\tau t_n,\dots).
\label{scalingD}
\eea
The presence of  this symmetry provides some selection rule for the terms of equations
of the hierarchy,
forbidding terms with higher symmetries (dispersion), therefore we call it dispersionless
scaling. If we consider the KP equation, only the terms corresponding to the dispersionless
KP equation possess this symmetry.
\subsection{Graded scaling}
Let us take $F^1=L$, $F^2=M$.
The  Orlov-Schulman flow written in terms 
of `plus' vector fields (\ref{OS+}) reads
\bea
&&
\frac{\partial}{\partial \tau}
\begin{pmatrix}
L\\
M
\end{pmatrix}=
\left
(\left(
\frac{L M_p}{\{L,M\}}\right)_+
{\partial_x}
-\left(\frac{L M_x}{\{L,M\}}\right)_+
{\partial_p}\right)
\begin{pmatrix}
L\\
M
\end{pmatrix}
+
\begin{pmatrix}
L\\
0
\end{pmatrix}.
\label{Scaling+}
\eea 
Using the identities
\beaa
\left(
\frac{LM_p}{\{L,M\}}\right)_+
=
\left(
\frac{L\p_p\sum t_n L^n}{\{L,M\}}\right)_+,
\quad
\left(
\frac{L M_x}{\{L,M\}}\right)_+
=
\left(
\frac{L\p_x\sum t_n L^n}{\{L,M\}}\right)_+
\eeaa
we rerite flow (\ref{Scaling+}) as
\beaa
\frac{\partial}{\partial \tau}L=L +\left(-p\p_p+ \sum_{n=1}n t_{n} \p_{n}\right) L   ,
\quad
\frac{\partial}{\partial \tau}M=\left(-p\p_p+ \sum_{n=1}n t_{n} \p_{n}\right) M,
\eeaa
and for the functions $u$, $v$ we get
\beaa
\frac{\partial}{\partial \tau}u=\left(\sum_{n=1}n t_{n} \p_{n}\right) u +  2u  ,
\quad
\frac{\partial}{\partial \tau}v=\left(-p\p_p+ \sum_{n=1}n t_{n} \p_{n}\right) v +v.
\eeaa
In explicit form,
\bea
&&
u(\tau)=\e^{2\tau} u(x,\e^\tau y , \e^{2\tau} t,\dots, \e^{n\tau} t_n,\dots),
\nn\\
&&
v(\tau)=\e^{\tau} v(x,\e^\tau y , \e^{2\tau} t,\dots, \e^{n\tau} t_n,\dots).
\label{scaling1}
\eea
This symmetry gives some graded homogeneity condition for the terms of equations of the
hierarchy.
\subsection{KP-type scaling} 
Let us take $F^1=1$, $F^2=LM$. 
This example represents 
a combination of generators of two previous examples.
The Orlov-Schulman flow
can be written as
\beaa
&&
\frac{\partial}{\partial \tau}L=L +\left(-p\p_p+ \sum_{n=0}(n+1) t_{n} \p_{n}\right) L   ,
\\
&&
\frac{\partial}{\partial \tau}M=- M + \left(-p\p_p+ \sum_{n=0}(n+1) t_{n} \p_{n}\right) M,
\eeaa
and for functions $u$, $v$ we get
\bea
&&
u(\tau)=\e^{2\tau} u(\e^\tau x,\e^{2\tau} y , \e^{3\tau} t,\dots, \e^{(n+1)\tau} t_n,\dots),
\nn\\
&&
v(\tau)=v(\e^\tau x,\e^{2\tau} y , \e^{3\tau} t,\dots, \e^{(n+1)\tau} t_n,\dots).
\label{scaling2}
\eea
For the reduction $v=0$ (in terms of the hierarchy ${\{L,M\}}=1$)
this symmetry coincides with the scaling symmetry possessed
by the KP hierarchy (with dispersion), thus we call it KP-type scaling.
\section{Orlov-Schulman symmetries in terms of the Riemann-Hilbert problem}
First we briefly describe the dressing scheme for the Manakov-Santini hierarchy,
following the work \cite{LVB09}.
A dressing scheme for the Manakov-Santini hierarchy can be formulated
in terms of two-component nonlinear Riemann-Hilbert problem on the unit circle $S$
in the complex plane of the variable $p$,
\bea
L_\text{out}=R_1(L_\text{in},M_\text{in}),
\nn\\
M_\text{out}=R_2(L_\text{in},M_\text{in}),
\label{RiemannMS}
\eea
where the functions 
$L_\text{in}(p,\mathbf{t})$, $M_\text{in}(p,\mathbf{t})$ 
are analytic inside the unit circle,
the functions $L_\text{out}(p,\mathbf{t})$, $M_\text{out}(p,\mathbf{t})$ 
are analytic outside the
unit circle and have an expansion of the form (\ref{form01}), (\ref{form1}) at infinity.
The functions $R_1$, $R_2$ are suggested to define a complex
diffeomorphism $\mathbf{{R}}\in\text{Diff(2)}$, and we call them
the dressing data. It is straightforward to demonstrate that the problem
(\ref{RiemannMS}) implies the analyticity of the differential form
$$
\Omega_0=\frac{\d L\wedge \d M}{\{L,M\}}
$$
(where the independent variables of the
differential include all the times $\mathbf{t}$ and $p$)
in the complex plane and the generating relation
\bea
\left(\frac{\d L\wedge \d M}{\{L,M\}}\right)_-=0,
\label{analyticity0}
\eea
thus defining
a solution of the Manakov-Santini hierarchy.

Let us consider in more detail, how the dynamics is introduced in terms of the dressing scheme.
Riemann-Hilbert problem (\ref{RiemannMS})
can be symbolically written in the form
\be
(L_\text{out},M_\text{out})=\mathbf{R}(L_\text{in},M_\text{in}),
\label{RiemannMSbis}
\ee
where $\mathbf{R}$ is a diffeomorphism, $\mathbf{R} \in \text{Diff(2)}$.
A natural way to introduce a dynamics is to define evolution of the dressing data
(diffeomorphism $\mathbf{R}$). Starting from one-parametric group of diffeomorphisms
$\mathbf{f}_\tau$, connected with some vector field $\hat {\mathcal{V}}$,
\bea
\hat {\mathcal{V}}={\mathcal{V}}^1(L,M)\p_L+{\mathcal{V}}^2(L,M)\p_M,
\label{vectorfield}
\eea
we define
\bea
\mathbf{R}(\tau)=\mathbf{f_\tau}^{-1}\circ\mathbf{R}_0\circ\mathbf{f_\tau}
\label{evolution}
\eea
The times of the hierarchy $t_n$ correspond to commuting vector fields $\hat {\mathcal{V}}^n=L^n \p_M$,
for which diffeomorphisms can be found explicitly,
\beaa
L \rightarrow L,\quad M \rightarrow M + \sum_{n=0}^{\infty}t_n L^n,
\eeaa
and we arrive to Riemann-Hilbert problem of the form (\ref{RiemannMS}), where 
diffeomorphism $\mathbf{R}$ is independent of times, and dynamics is defined by the
`singular terms'  of $M$ (\ref{form1}), in some similarity with the Baker-Akhiezer function.
To this Riemann-Hilbert problem we can consistenly add evolution of the dressing data of the form
(\ref{evolution}), corresponding to some vector field (\ref{vectorfield}).  The flows corresponding to
non-commuting vector fields do not commute. These flows represent Orlov-Schulmann symmetries
of the MS hierarchy in terms of the dressing scheme.

To obtain  Lax-Sato equations (\ref{OS+}) for the Orlov-Schulmann flows, one needs to add some extra
relations to the generating equation (\ref{analyticity0}). They can be obtained by modifying the differentials
\beaa
&&
\d L\rightarrow \d L+(\p_\tau L + (\hat {\mathcal{V}} L))\d \tau,
\\
&&
\d M\rightarrow \d M + (\p_\tau M + (\hat {\mathcal{V}} M))\d \tau,
\eeaa
then generating equation (\ref{analyticity0}) provides all the necessary extra relations.

Taking commuting vector fields $\hat {\mathcal{V}}^n=L^n \p_M$, we reproduce dynamics corresponding 
to the times of the hierarchy for $L$ and nonsingular part of  $M$.
Vector field of the form $\hat {\mathcal{V}}={\mathcal{V}}^2(L,M)\p_M$ leads to Lax-Sato equation  (\ref{OS+})
with $F^1={\mathcal{V}}^2(L,M)$, $F^2=L$, and for vector field 
$\hat {\mathcal{V}}={\mathcal{V}}^1(L,M)\p_L$ we have
$F^1=-{\mathcal{V}}^1(L,M)$, $F^2=M$. For a general vector field (\ref{vectorfield}) we have  
linear combination of symmetries  (\ref{OS}) of the form
\bea
&&
\frac{\partial}{\partial \tau}\begin{pmatrix}
L\\
M
\end{pmatrix}=
\left(\left(
\frac{{\mathcal{V}}^1 M_p -{\mathcal{V}}^2 L_p}{\{L,M\}}\right)_-
{\partial_x}
-\left(\frac{{\mathcal{V}}^1 M_x-{\mathcal{V}}^2 L_x}{\{L,M\}}\right)_-
{\partial_p}\right)\begin{pmatrix}
L\\
M
\end{pmatrix}\qquad
\label{OSV}
\eea 
or, writing in terms of infinitesimal symmetries,
\bea
&&
\delta_{\hat {\mathcal{V}}}\begin{pmatrix}
L\\
M
\end{pmatrix}
\nn
\\&&\quad
=
\left(\left(
\frac{1}{\{L,M\}}
{\begin{vmatrix}
\hat {\mathcal{V}} L& \hat {\mathcal{V}} M\\
L_p & M_p
\end{vmatrix}}
\right)_-\p_x
-
\left(
\frac{1}{\{L,M\}}
{\begin{vmatrix}
\hat {\mathcal{V}} L& \hat {\mathcal{V}} M\\
L_x & M_x
\end{vmatrix}}
\right)_-
{\partial_p}\right)\begin{pmatrix}
L\\
M
\end{pmatrix}\qquad
\label{OSVinf}
\eea 
Commutator of infinitesimal symmetries (\ref{OSVinf}) is a symmetry, corresponding to a commutator
of respective vector fields (\ref{vectorfield}),
\beaa
[\delta_{\hat {\mathcal{V}}_1},\delta_{\hat {\mathcal{V}}_2}]=
\delta_{[\hat {\mathcal{V}}_1,\hat {\mathcal{V}}_1]},
\eeaa
and the linear map $\hat {\mathcal{V}}\mapsto \delta_{\hat {\mathcal{V}}}$ defines a 
homomorphism of a  Lie algebra of two-dimensional vector fields into the algebra
of symmetries of the MS hierarchy (compare \cite{Takasaki92,Takasaki95}).
 
For Hamiltonian vector field $\hat {\mathcal{V}}=H_M(L,M)\p_L - H_L(L,M)\p_M$
we have
$F^1=1$, $F^2=H(L,M)$. The evolution (\ref{evolution}) for this vector field
preserves the Jacobian of $\mathbf{R}$, thus it is compatible with the Hamiltonian reduction 
defining the dKP hierarchy (${\{L,M\}}=1$), and formula (\ref{OSVinf}) transforms to infinitesimal
Orlov-Schulman symmetries 
for the dKP hierarchy, see \cite{Takasaki92,Takasaki95}.
 
For the elementary examples of Orlov-Schulman symmetries that 
we considered above (Galilean transformation and scalings) 
vector fields $\hat {\mathcal{V}}$ have zero
or constant divergence, and they are compatible with the Hamiltonian reduction.
\subsection*{Funding}
{This research was performed in the framework 
of the State Assignment of the Ministry of Science and
Higher Education of the Russian Federation
(topic FFWR-2024-0012
``Quantum field theory, string theory and mathematical physics'')}
\subsection*{Conflict of interest}
The author of this work declares that he has no conflicts of interest.

\end{document}